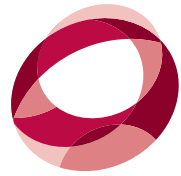

# Next Generation Robotics

Version 1

*Editorial team:* Henrik I Christensen, Allison Okamura, Maja Mataric, Vijay Kumar, Greg Hager, Howie Choset

*Significant input from:* Peter Allen, Aaron Ames, Brenna Argall, Ruzena Bajcsy, Calin Belta, Mark Campbell, Dieter Fox, Bobby Gregg, SK Gupta, Martial Hebert, John Hollerbach, Lydia Kavraki, Hadas Kress-Gazit, James Kuffner, John Lizzi, Mac Schwager, Mark Spong, Yu Sun, Reid Simmons, Lynn Parker, Dmitry Berenson, Nikos Papanikolopoulos, Missy Cummings, Tim Bretl, Julie Shah, Seth Hutchinson, Jana Kosecka, Conor Walsh, Jaydev Desai, Mohan Trivedi, and Daniela Rus

# I. Introduction

The National Robotics Initiative (NRI) was launched 2011 and is about to celebrate its 5 year anniversary. In parallel with the NRI, the robotics community, with support from the Computing Community Consortium, engaged in a series of road mapping exercises. The first version of the roadmap appeared in September 2009; a second updated version appeared in 2013. While not directly aligned with the NRI, these road-mapping documents have provided both a useful charting of the robotics research space, as well as a metric by which to measure progress.

This report sets forth a perspective of progress in robotics over the past five years, and provides a set of recommendations for the future. The NRI has in its formulation a strong emphasis on co-robot, i.e., robots that work directly with people. An obvious question is if this should continue to be the focus going forward? To try to assess what are the main trends, what has happened the last 5 years and what may be promising directions for the future a small CCC sponsored study was launched to have two workshops, one in Washington DC (March 5th, 2016) and another in San Francisco, CA (March 11th, 2016). In this report we brief summarize some of the main discussions and observations from those workshops.

We will present a variety of background information in Section 2, and outline various issues related to progress over the last 5 years in Section 3. In Section 4 we will outline a number of opportunities for moving forward. Finally, we will summarize the main points in Section 5.

# 2. Background

As mentioned earlier the National Robotics Initiative (NRI) was launched September 2011 and has had five rounds of call for proposals. The NRI is coordinated by NSF but with active involvement and support from NSF,

NASA, USDA, NIH, the Department of Defense (DOD), the U.S. Department of Energy (DOE)[1] and OSTP. The stated goal of the National Robotics Initiative is "to accelerate the development and use of robots that work beside or cooperatively with people in the United States."

The basic research themes in the NRI solicitation include:

◗ Sensing and perception

◗ Design and materials

◗ Modeling and analysis of co-robots

◗ Human-robot interaction

◗ Planning and control

There is also an emphasis on STEM education through robotics, as well as on research to understand long-term social, behavioral, and economic implications of co-robots.

In addition to the basic research focus, the participation of mission-oriented federal agencies brings a broader perspective to the NRI. There are new applied research and development themes as well as multi-faceted collaborative efforts in diverse application sectors including agriculture, defense, medicine and space.

The first year of funding (FY 12) funded 61 proposals at a total of over $40M/year. Since then, more than 200 proposals have been sponsored at a total of more than $150m by the partner agencies. The majority of the sponsored projects are still underway. A few projects have graduated to the i-Corp program for translation into start-up companies or been adopted by corporations such as Marlin Wire, P&G, BMW, and Intuitive Surgical.

Two workshops have been organized during the last year to consider issues related to the National Robotics Initiative. One was directed at the relation between Cyber Physical Systems (CPS), the NRI and the need for systems with a higher degree of Autonomy (Future

---

Directions in Cyber-Physical Systems, Robotics, and Autonomy, NSF Workshop, Sept 2015). Another was directed at the formulation of a Synthetic Science of Physical Intelligence organized by CCC and taking place at UPENN October 19-20, 2015. In a closely related activity, the Computing Community Consortium initiated a series of white papers on the "Science of Autonomy" in summer 2015.[2]

There is clearly a need to consider how different programs related to the integration of physical interaction, perception, and artificial intelligence can be coordinated to ensure that USA remain at the forefront of the research area and provides both the bets R&D but also human resources for the industry. This was called for in the recent review of the Networking and Information Technology R&D (NITRD) program[3]. Subsequent to this report, it is encouraging to see that a new Working Group has been setup under NITRD to support a new Robotics and Intelligent Systems Program Component Area.

## 2.1 NRI Drivers

One of the main drivers of the NRI is the potential to improve economic productivity and the quality of life of the ordinary citizen through robotic technology. Robotic technology has had a huge impact in areas where we can now do new things we could not do before – the technology has increased existing human capabilities. Some examples of this include robotic surgery systems, autonomous cars, and "smart" agriculture that increases yields and reduces waste of water and fertilizer.

Robotic capabilities have improved greatly over the past few years, in part due to the expanded NRI effort, and advances in mobility, manipulation and sensing/mapping are making inroads into many markets and products that can benefit from these capabilities. Space has been a prime example domain for robotics, but undersea applications are also growing, ranging from aquaculture, to the repair and maintenance of pipelines/cables.

Another important application area is disaster prevention and recovery. Robots can prevent disasters; two examples of rapidly growing industries are unmanned aerial systems for inspection of critical infrastructure to prevent incidents, and underwater robots for detection of smuggling and terrorist activities around major ports. Robots can save lives and reduce the economic consequences of disasters as seen in over 20 incidents in the USA including robots capping the leak at the BP Deepwater Horizon Oil Spill.

Robotic technology has also had a major impact on our quality of life. Home health care, mobility, wellness and well-being are being positively impacted by assistive robotics, human-robot interaction, advanced prosthetics, and smart sensing, all areas that are central to the NRI. The emergence of "Smart Cities" and Internet of Things (IOT) initiatives led by private industry is supported by new sensing and robotic technologies coupled with advanced networked software, all components of NRI research.

Finally, Robotics can be seen as a tool for not just enhancing but potentially revolutionizing K-12 STEM education, both formal and informal, in order to train a competitive 21st century US workforce, lower the digital divide, and bring more gender and ethnic balance to the STEM workforce. In this context, social robots can boost the confidence and self-esteem of children from all socio-economic backgrounds, potentially even in families that may not appreciate the importance of STEM education, or education of any kind.

## 2.2 NRI Impacts

One of the major impacts of NRI funded research is that it forced many researchers to look beyond their own limited, niche domains and expand their research horizons by collaborating with other researchers to build new systems and applications that involved both humans and robotics (co-robotics). Many of the PIs and students who have been supported by NRI are researchers from disciplines outside of the traditional core robotics areas. These collaborations have been quite fruitful in creating a much broader and inclusive set of domains for robotics research and applications. Central to this objective is putting researchers into real environments, populated with humans and physical robots.

Another major impact is the open-sourcing of robotics hardware and software. This trend continues to accelerate with positive benefits accruing. Before NRI, it was quite difficult and expensive to build and equip a laboratory focusing on robotics. That cost has been driven down by the emergence of inexpensive and replicable hardware (arms, vehicles, humanoids, sensors etc.) along with open-source libraries devoted to many of the most useful robotic algorithms (planning, control, imaging etc.), all configured to run under the open-source Robotic Operating System (ROS). ROS itself is supported by NRI, and most NRI projects are developing software that can be open-sourced as well. This effect has streamlined and shortened the learning and implementation curves for most robotics researchers while making access simpler for new entrants into the field. Building a complex robotics system, which used to take years, can now be accomplished in months instead. Further, large databases of objects, environments, and physical components have been created and re-used across the community, supporting the trend in large cloud-based computing resources available to all.

A further impact is the benefit that robotics brings to STEM education. Robotics can make STEM courses come alive with engaging physical robots that students can build, program and from which they can learn directly. National Robotics Week, celebrated every April, has blossomed into an effective and far-reaching way to spur students into the robotics and other STEM fields. NRI supported researchers and students are at the front lines of presenting forums, demos and open houses that effectively let the public know about the growth and potential of robotics. STEM education has become a strong national priority. Employers are desperately looking to fill new jobs with qualified STEM graduates. In the robotics sector alone, large industrial organizations such as Apple, Google, Amazon, Uber, Tesla are looking to hire many new robotics engineers, many of whom are coming out of NRI funded programs.

Another impact is that robotics-based STEM training can be more appealing to underrepresented groups such as women, helping to create better gender and socio-economic balance in our country. The appeal of the NRI program has also crossed Federal funding agency boundaries, with participation from NIH, DOD, DOE, USDA and NASA. This helps to further grow the field as robotics enters more and more aspects of our society.

One of the most important metrics for the NRI program is the explosive growth of robotics research across the globe. As interest in robotics increases, there is now a burgeoning and strong community of roboticists. This can be easily measured by:

1. Increased attendance and submissions of papers at the major robotics conferences. At the most recent IROS conference in Hamburg (10/15) there were 2134 contributed paper submissions,45 sessions in 15 parallel tracks, 51 accepted Workshop and Tutorial submissions, 72 accepted Late Breaking Poster papers, 6 plenary and 9 keynote talks, and over 2500 registrants. At ICRA 2015 in Seattle there were over 3000 attendees (an ICRA record). Highlighting the conference were 940 accepted technical papers (out of 2275 submissions) presented over 3 days in 10 parallel tracks, representing authors from over 40 countries. There were also over 1400 attendees (another ICRA record) participating in 42 workshops and tutorials. The conference also highlighted the increasing role of women in robotics, with a General and Program Committee that was entirely female.

2. Development of a wide range of offshoot conferences and workshops focused on robotics topics, as diverse as UAV's, Surgical Robotics, Planning and Control, Humanoids, Disaster and Safety, Ubiquitous robots, and Benchmarking. These are just a few examples from conferences coming up in next few months). Similarly, there are many new academic journals devoted to robotics (e.g. IEEE Robotics and Automation Letters, Soft Robotics, Robots and Biomimetics, Journal of Robotics, Networking and Artificial Life, Journal of Human-Robot Interaction).

3. In academia, evidence of this impact can be seen in (a) increased student enrollment in robotics courses at the undergraduate and graduate levels, (b) new and growing robotics departments, centers, and programs at the undergraduate, master's and doctoral levels, and (c) faculty hiring in robotics has also significantly increased due to the factors above.

4. Private industry is equally interested in robotics. The number of jobs for students continues to grow showing the interest and need for trained roboticists in the industrial sector. Marquee companies like Uber, Google, Amazon, Apple, and Tesla are all looking for graduates trained in robotics, as are the numerous startups that have been created over the last few years. While some of this has been disruptive for academic research (e.g., because of faculty being recruited to start ups), the overall impact on the field has been positive.

5. Open source platforms, databases, code repositories have proliferated. Industrial manufacturers of robots are now almost required to provide an open source ROS interface to their products for them to be successful. GITHUB and ROS repositories now allow new players easy access to developing new robots and capabilities.

6. Hardware has also become less expensive as more companies are building it. This reduced hardware platform cost has also reduced entry barriers for those wanting to do robotics research.

These metrics show that the NRI has been an enabler and catalyst for the growth of robotics as both a scientific discipline and economic force. However, this is only the tip of the iceberg in terms of what the US needs to train and employ a 21st century STEM workforce and to remain competitive internationally.

## 3. Recent Progress

Over the last 5 years we have seen tremendous progress both in terms of new applications of robotics and the component sciences. We will briefly summarize some of the examples of such progress in this section.

It is important to recognize upfront that robotics is still a very hard problem. While there are a number of technology demonstrations in robotics that suggest that they are becoming mature, it is also clear that many of these solutions only work under tightly constrained conditions and, are at best "demos". The recent Defense Advanced Research Projects Agency (DARPA) Robotics Challenge serves to highlight many of the open problems in robotics in addition to underscoring the tremendous potential of this field.

We may be able to drive a 1 ton vehicle autonomously for 1.5M miles[4], but the technology relies on detailed maps and is not robust to bad weather. In addition, we are not even close to understanding (or managing) the complex social interactions that occur between car and driver and between cars.

We might be able to design neural networks to learn the correct features to beat the world champion at Go, but that same neural network cannot beat a 5 year old at tic-tac-toe.

Industrial robots routinely pick up and manipulate parts in a structured industrial setting, but the lack the dexterity of a 3-year old playing with Lego blocks.

A lot of progress has been achieved over the last 5 years, as outlined below, but it is far from a solved problem.

---

[4] https://www.google.com/×selfdrivingcar/

## 3.1. Actuation / Materials

In actuation we have seen major progress both in terms of miniaturization and utilization of new materials. One such example is the development of micro-sized flying vehicles[5], which has required research on active materials, on visual processing, and systems integration. This is a great example of how multi-disciplinary research is required to generate a leap in performance. New MEMS and Material Science has also allowed design of new types of grasping systems and soft robots . A number of studies have demonstrated that robotics is not just about integrating existing components, but also the multi-disciplinary discovery of new methods for design of systems that have superior performance. The joint research on walking between UPENN, CMU and GT is another great example of such work.

## 3.2. Big Data / Analytics

We have seen a tremendous growth in the availability of sensors for monitoring of processes over the last decade. In addition, we have seen exponential growth in the availability of computer power for data processing. The graph below illustrates how Graphical Processing Units (GPU) have emerged as desktop mini-computers computer signal/image processing.

Evolution in computing power for CPUs and GPUs over the last decade

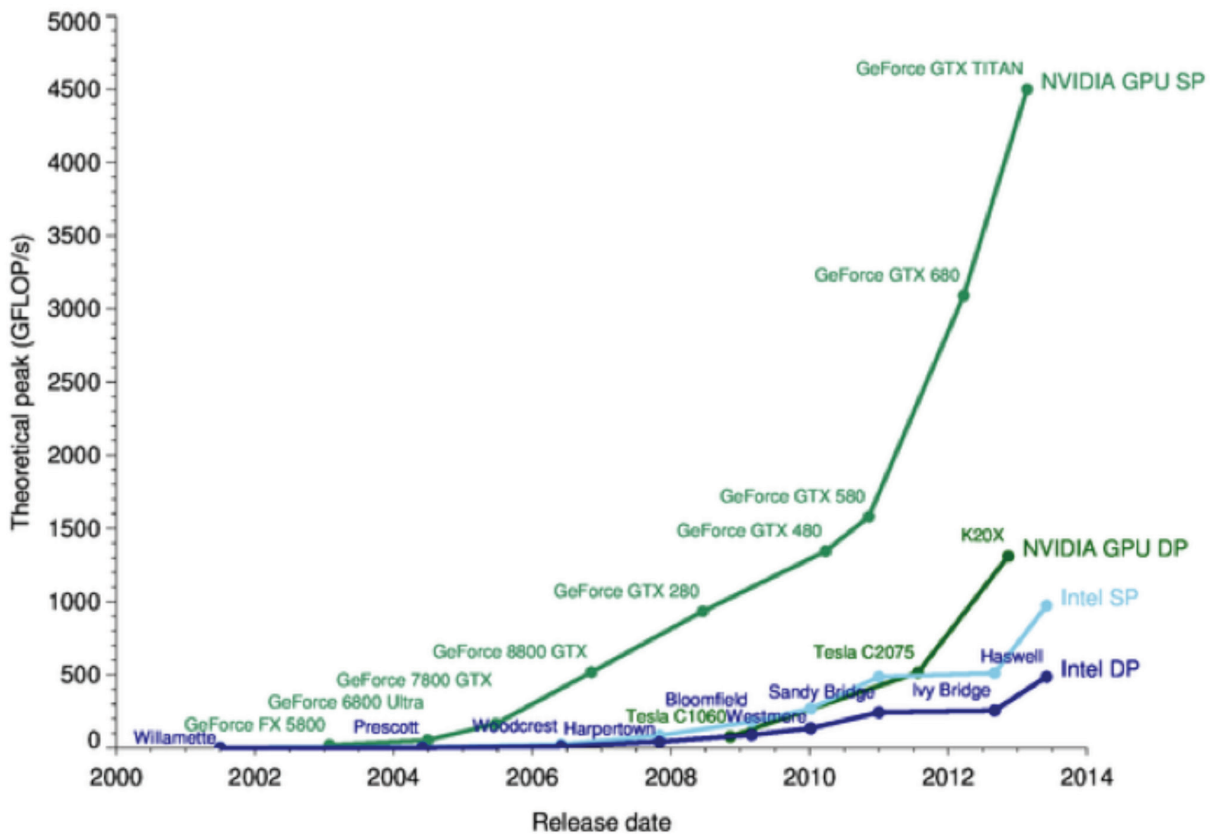

Evolution in computing power for CPUs and GPUs over the last decade[7]

The amount of data available per person has double every 40 month since 1980. Year 2012 the amount of data generated every day was 2.12 exabyte ($2.1*10^{18}$). It is anticipated that the big winner in terms of utilization of data will be in manufacturing due to improved process monitoring and optimization of the supply chain[8].

The adoption of big data varies tremendously across sectors. The main drivers have been in finance and real-estate, whereas manufacturing/healthcare is just now starting to see real impact.

See (Lee, Bagheri, & Kao, 2015) for a discussion of recent progress on big data architectures for robotics and automation.

Big Data processing and the use of Graphical Processing Units (GPUs) has already revolutionized image processing. The area of machine learning termed deep learning[9] has facilitated a new level of performance in image based diagnostics and recognition, which has motivated companies such as Facebook, Google and Microsoft to make major investments in these technologies. It is important to recognize that there is an abundance of data and processing power but this far limited progress has been achieved on turning data into actionable information. The biggest challenge remains model-based data processing for monitoring and controlling tasks in real-time.

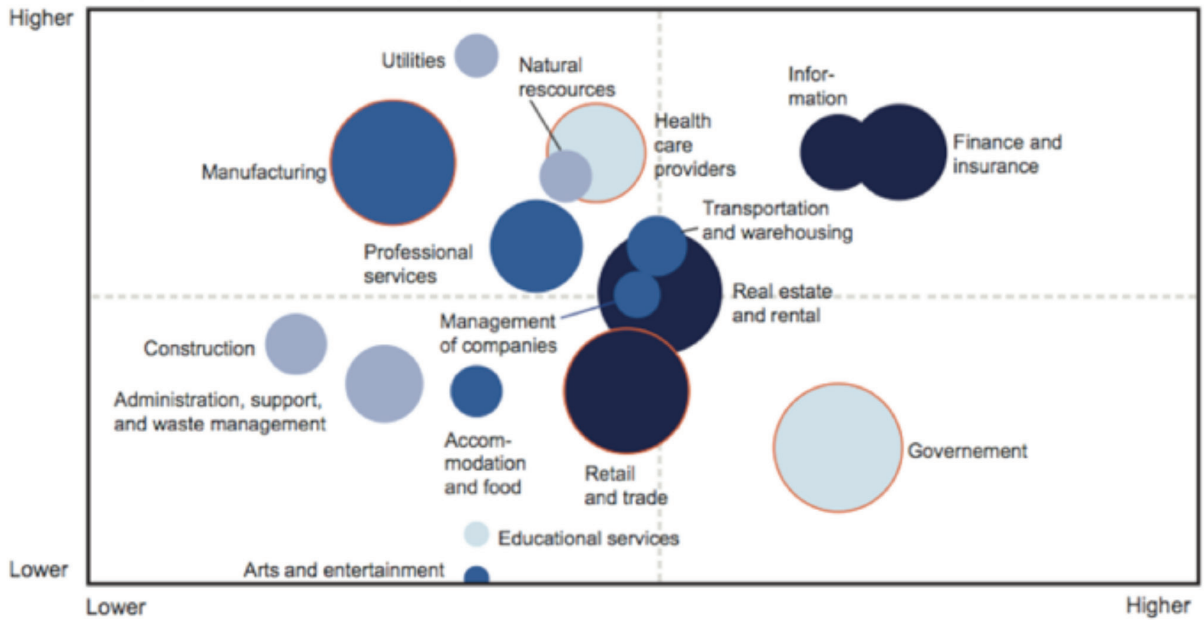

SOURCE: US Bureau of Economic Analysis; McKinsey Global Institute analysis

---

Several technologies of direct relevance are mapped out in the Gartner 2015 Hype Cycle shown below

It is encouraging to see 3D printing as a short-term technology, but it is interesting to see that intelligent robots are considered 5+ year away and so are Smart Advisors, Internet of Things and Digital Security. Nonetheless it does give an outside perspective on the maturity of different technologies.

## 3.3. Software Generation

Progress on software systems for automated planning, verification and code-generation has been significant over the last decade. Initial progress was driven by academic research but with limited complexity systems. Over the last few years, progress has been achieved through a number of major projects. The most well-known is probably the Adaptive Vehicle Make (AVM) program[10] sponsored by DARPA, where the objective is to manufacture a military vehicle directly from the engineering design files. The project has since then become part of the Digital Manufacturing NNMI institute[11], which has significant support from several major companies such as GE. Several projects across the world, but very much dominated by the automotive sector, are driving automatic generation of software for manufacturing processes. As the project variation, while potentially large, is deterministic it is possible to design a process that is relatively deterministic. The NNMI institute on Digital Manufacturing has yet to release a technology roadmap for general industries.

In Europe there are a number of major efforts underway as part of the Horizon 2020 program. Again most of the programs are driven forward by the automotive industry. The vision for Europe has been proposed by the HYCON network[12] and the follow-up CPSoS[13] support action. More recently the big driver has

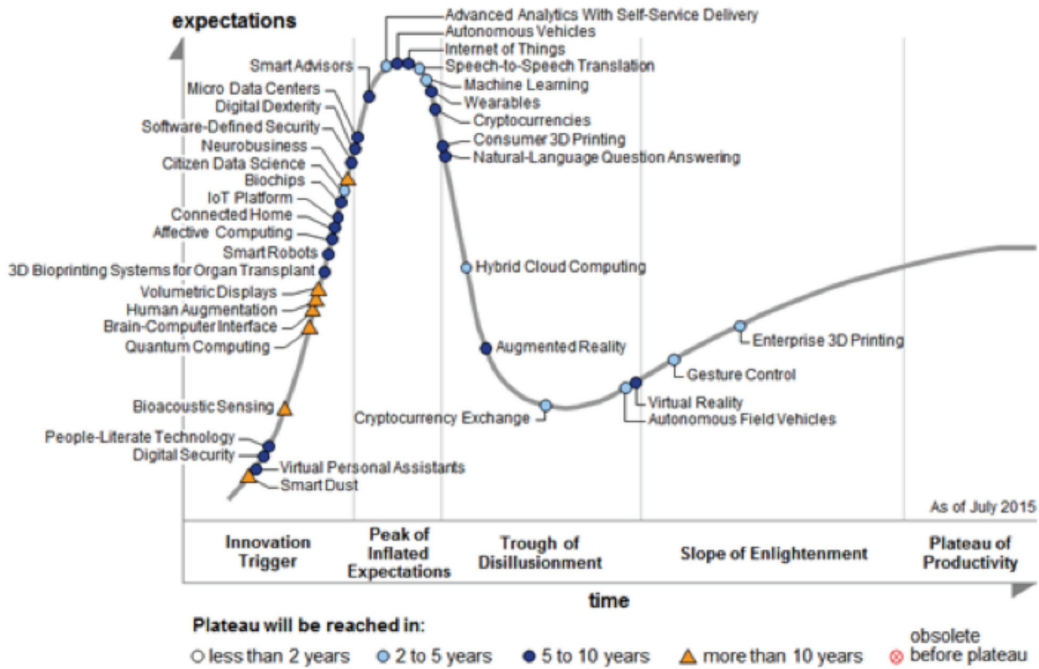

Gartner 2015 Hype Cycle for Emerging Technologies

been the Horizon 2020 – Factory of the Future program, which has its emphasis across design, manufacturing, deployment and maintenance. The program is funded at $1.2B over 2014-2020. The roadmap is available online[14]. So far limited emphasis was been devoted to software generation for low-rate manufacturing processes.

In the domain of robotics the Industrial Robot Operating System (ROS-I) eco-system has developed the systems MoveIt[15], which is a robot task-planning framework that allow automatic code-generation. The system is a first step towards automated code generation for robot systems. NIST has recently launched an effort to standardize a "simplified" robot language to allow automated task planning (using PDDL), automatic code generation and execution. The test cases are still relatively simple for cases such as kitting.

There is no doubt the tool suites are emerging for automated code generation from engineering design to task specification, to NC and/or robot program generation. The AVM program solved the complexity problem through use of standardized sub-assembles. There is a clear need for more efficient code generation and for methods to verify execution prior to use.

## 3.4. Collaborative Systems

Over the last few years we have seen tremendous progress on collaborative systems and human-safe robots. The progress easily seen in terms of new human-safe collaborative robot systems such as KUKA iiwa, Universal Robots, Rethink Baxter and Sawyer, Cyber Dyne systems, etc. Today the fastest growing market segment is collaborative robots which has a growth rate of 50% per year compared to traditional industrial robot systems that have an annual growth rate of 16%.

Equally important, we have seen tremendous progress on the design of user interfaces that allow easy / quick programming of robots for particular tasks. We have seen major progress and proliferation of groups that do research on collaborative systems both for software generation[16] and learning by demonstration[17].

## 3.5. Major application areas

Manufacturing has seen a major renaissance over the last 5 years. The sector continued see 12-18% per year growth and has recently reported the best robot sales numbers ever. About 40% of industrial robot sales are in manufacturing. Major new growth sectors has been re-shored electronics manufacturing and use of robots for supply chain and e-commerce. Online sales have grown more than 40% per year and resulted in major investments by companies such as Amazon, Target, and Walmart. The big drivers have been improved quality of products and increased agility. At the same time the new applications has unraveled a need for improved robot perception and handling of more complex object shapes.

For domestic robot applications we continue to see major growth in the basic robot navigation space with more than 10,000,000 units sold. An encouraging aspect is that these robots are starting to utilize Visual SLAM for the mapping and navigation. It is now possible to get a cell phone camera and pair it up with a cellphone processor for doing automated mapping in dynamic environments such as a regular house at a cost of less than $100. This progress is opening up for a large variety of new applications.

For robotic surgery more than 600,000 minimally invasive procedures are performed each year by the da Vinci Surgical System, and more than 3 million procedures have been performed since 2000. Research into medical robotics has enabled improved imaging integration, improved procedures, improved team training and new opportunities for integration of pre-operative

---

planning. In most cases a minimally invasive procure allow people to return to work/home much quicker, the risk of complication is reduced and the operating time is reduced to free up capacity at hospitals.

Driverless cars have driven more than 6 million miles. It is already legal to operate driverless cars in 4 states (CA, NV, MI and FL). Already today major car companies provide level 2 autonomy in their products. This includes lane keeping, active breaking, traffic sign registration, car-to-car communication, automatic (and remote) parking, etc. The expectation is that most of the major providers will have products on the market within 3-4 years. Much of the progress has been enabled by improved sensors (Camera[18] & Radars), availability of new computing platforms (NVIDEA and Intel) and use of deep learning.

Already today 40% of the pilots entering the military for pilot training become drone operators and there is tremendous growth in utilization of unmanned aerial systems (UAS) for applications such as crop monitoring, construction site verification, mineral exploration, disaster mitigation, and site planning. The technology is available today to allow for autonomous delivery in supply chain applications. The main limitation is in the legal framework to enable broader introduction of such vehicles into the national airspace. Other technical limitations to these systems today are in terms of payload, battery time and the sensory suite that can be accommodated on a platform.

## 3.6. Academic Growth

Over the last 5 years there has been major growth in new academic program and the organization of new academic units. Several universities have setup new research centers in robotics (UMICH, ASU, Oregon State, UCSD, …) and in addition a number of new educational programs have emerged both at the undergraduate and graduate levels (WPI (B.Sc. degree), CMU (B.Sc. minor), UPENN (M.Sc.), …)

In addition, the number of new academic positions has also grown very significantly. Last year there were more than 50 openings for robotics faculty, which is a radical change from a few years ago. The NRI in some sense has provided validation that robotics is major subject across a variety of different subjects.

## 4. Moving forward

The field has seen tremendous progress over the last 5-10 years. However, robotics is far from a solved problem and the penetration into most domains is still at its infancy. There is a continuous dialog about the bets way to organize research. Should research be defined to try to solve "moon shots" as we saw with NASA 50 years ago or it is better to define research in terms of core topics that should be addressed to enable a broader set of applications? Recently there has been a push for definition of moon shots. The NASA moon mission has an estimated cost of $5.2B[19] at the time. The mission had a broad set of societal benefits from new materials to control and aeronautics. However it is less clear that smaller programs would have similar impact.

## 4.1 Moonshots

As part of the workshops several potential "moon shot" candidates were defined. Some of them are briefly summarized below.

Driverless cars have to the potential to significantly reduce the number of traffic casualties. Today more than 33,000 people are kill in US traffic and the number is close to 1.4 million world wide[20]. Reducing this number by an order of magnitude would have a tremendous economic and societal impact. According to NHTSA the cost of road accidents in 2010 was $1 trillion for that year along in terms of loss of productivity and lives. Design of driverless vehicles requires further progress

---

on sensors, sensory fusion, active control, vehicle to vehicle communication, fleet management and user interfaces for non-expert users.

All the western societies are experiencing significant changes in demographics. An interesting "moon shot" could be design of assistive robots that would allow people to retain the quality of life (with respect to aging) for another 10 years. This would reduce the cost of healthcare significantly. Over time we reduce our mobility and mental capabilities. Alzheimer and other memory deficiencies have a significant impact. There is a rich set of opportunities across mobility support, daily functions such as getting out of bed, getting a shower, getting dressed, preparing a meal, and getting reminders about medication and exercise.

A related challenge proposed was eliminating disability. The proposed mission would be to eliminate disability to a degree where the American Disabilities Act no longer would make sense. How can we design a spectrum of assistive devices that would allow all people with disabilities to be 100% participants irrespective of their disability? This has interesting consequences for design of brain computer interfaces, exoskeletons, prosthetic devices, etc. This would be even more interesting if the devices were design to adapt over time as the user and their environment change over time.

A fourth area would be production of food. We are quickly running short on food and it has to be more efficient to produce food and put it in the hands of people worldwide. One opportunity could be production of food in half the amount of space and with the use of half the amount of water. This would make food production more economically viable or we could make twice as much food without any increase in cost. In food manufacturing there are enormous opportunities for quality control, increased productively and reduction of cost.

## 4.2 Application Drivers

An alternative approach for definition of a research program is through a direct consideration of business drivers. The clear business drivers include

❱ 1 off manufacturing

❱ Automated Software Generation

❱ Service robots for daily assistance

❱ Field Robots for Assistance in Disaster Recovery

Traditionally, production systems have been used for mass manufacturing. This is no longer a valid model of manufacturing. Consumer products are made in many varieties. As an example the AUDI A3 is made is 6 million different configurations. The personalization challenges automated manufacturing. In automotive manufacturing the plate shop, welding of the chassis and the paint for the basis chassis is fully automated however the final assembly has not been automated due to lack of an ability to customize processes to manage millions of variations in the process. How can we design robot systems that allow handling of a very significantly set of variations? This requires flexibility in end-effector, sensor based tracking objects, online changes in software configurations and methods for automated. The change of mass manufacturing to agile 1-off manufacturing will challenge programming, supply chain management, sensing for assembly, etc.

The process of programming robot systems is considered labor intensive. Many different aspects have to be considered as part of the design and implementation. In manufacturing the rule of thumb is that the cost of a system is 30% the robot, 20% auxiliary hardware and 50% of the cost is software. For broader adoption of systems and for quick adaptation of robot systems to new applications there is a need to automate the generation of software. How can we design systems such that domain knowledge is used to generate the software with minimum or no human intervention? The AVM project from DARPA made some initial progress for system configuration, but not clear it closes the loop for real-time execution. Still a fair number of challenges to address to make this manufacturing ready.

Service robotics for daily assistance has a tremendous potential given the changes in demographics. Soon 50% of the population will be above 40. With age comes a number of challenges such as reduced sight, hearing, mobility, memory, and dexterity. Robots offer an opportunity to address some of the needs such as medical reminder, exercise assistance, transportation of material, personal hygiene, ... The average cost of nursing assistance in a home is $10,000 / year. Design of a home robot that is economically viable and providing major assistance is interesting but also a major challenge. So far no-one has managed to deliver systems that truly deliver in terms of cost, robustness and performance.

After the Fukushima disaster there was a pickup in projects directed at assistance in emergency situation and management of nuclear risks. Unfortunately, so far little real progress has been achieved. Robot systems has been used to construct the sarcophagus for Chernobyl, and similar robot systems are used to clean up the reactor 3 at Fukushima. The cost and time to deploy such systems is very significant. For disaster management this is a need to survey the impact of an incident, to provide immediate assistance to reduce the impact and a need for long-term recovery. Mixtures of construction systems, unmanned aerial vehicles and ground robots have been deployed. DOE has started to consider use of robots for management of the nuclear waste already present at a number of storage facilities and a separate roadmap is due by the summer of 2016. An important aspect here is the need to team up with domain experts to ensure that real solutions are provided which provides real relief.

## 4.3 Research Evolution

As is already noted above, there has been an astonishing growth in the breadth and maturity of a variety of robotics-enabling technologies, as well as substantial progress on several major research themes of the robotics roadmap and the NRI. Examples of technologies that are reaching a new level of capability include:

1)  Perception – particularly video and depth image interpretation – due to advances in machine learning, data mining, and the availability of large data sets for training of machine vision systems. This has also been driven by the introduction of several low-cost video-plus-range (RGBD) imaging systems. As a result, we are seeing, for the first time, robust and wide-spread use of computer vision to guide vehicles, to support manipulation, and to enable human-computer interaction.

2)  Machine learning – much of perception has been driven by advances in machine learning. We are also seeing more exploration of learning-based methods in robotics, although as we further discuss below, the application of learning for robotic systems is not yet as widespread as in other areas of AI-related research.

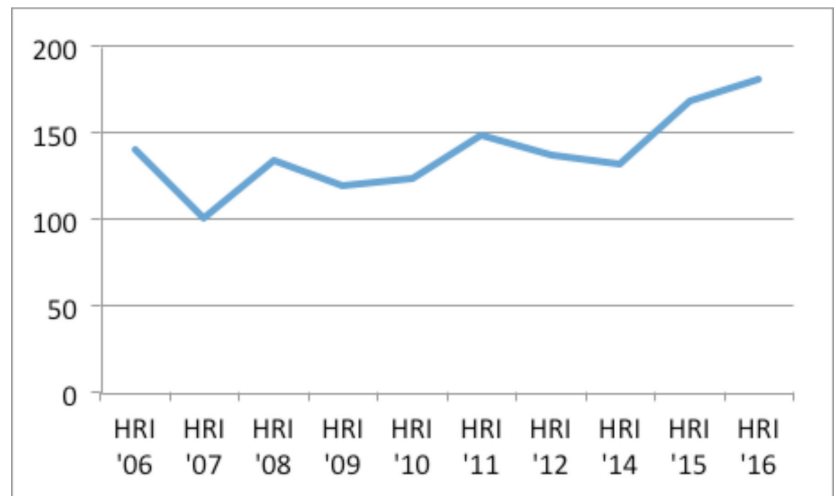

Submissions to the Human-Robot Interaction (HRI) conference over the past 10 years have risen nearly 50% in the past two years.

3) Human-robot interaction – the evolution of better platforms, better perception, and increasingly powerful software capabilities has supported a significant growth in the number of robotic systems that include some type of human-computer interaction component.

4) Low-cost hardware – it is now possible to purchase highly capable platforms of all types – ground-based, flying, manipulation, etc. – at very reasonable cost. This has accelerated the development of real-world systems and real-world experimentation.

5) Human-safe robots – the last five years has seen several human-safe robotic platforms fielded, as well as a growing acceptance of direct human-robot physical interaction as a "standard mode of operating."

6) Maturation of control, mapping, and planning – as with perception, increasingly powerful tools for control, localization and mapping, and robot planning are now widely available to the research community.

7) More accessible integrated systems – the amalgam of the above advances suggest that it far easier today to develop and test fully integrated robotic platforms than ever in the past.

## 4.3.1 Autonomy vs. Collaboration

A hallmark of the current NRI has been the focus on collaboration – creating systems that operate to complement or enhance human capabilities or productivity. A complement to collaboration is *autonomy*, which we define as a property of a system that is able to achieve a given goal independent of external (human) input while conforming to a set of rules or laws that define or constrain its behavior. The key point is that explicit execution rules are not (and cannot) be defined for every possible goal and every possible situation. For example, an autonomous car will take you to your destination (a goal) or park itself (another goal) while obeying the traffic laws and ensuring the safety of other cars and pedestrians. An autonomous tractor will till a field while avoiding ditches and fences and maintaining safety of the equipment and any human operators. An autonomous bricklaying system will build a wall in many different situations and with many different materials while ensuring the wall conforms to both building plans and building codes.[21] In short, a key difference is that autonomous systems must be able to act independently and intelligently in dynamic, uncertain, and unanticipated situations, but also it must be able to detect when its goals stand in conflict with the laws that govern its behavior, and it must have a way to "fail" gracefully in those situations.

Autonomy is in fact a key capability for collaborative systems – a collaborator must be able to operate independently, but with the "rules of engagement" for whatever the collaboration is. Despite what we see in the popular press, or the latest viral video, achieving this future vision is *emphatically not within the scope of today's technologies* – it requires substantial advances in both our technical and socio-technical understanding of the science of autonomy. It requires systems that are capable of receiving and carrying out natural language instruction at a relatively high level. It requires systems that can be physically capable in an environment that is unstructured and in situations that were never anticipated or tested. It requires systems that can co-exist with people, and be trusted, safe companions and co-workers.

The applications that demand some level of autonomous capability are wide-ranging and automated transportation (ground, water, and air), construction, agriculture, manufacturing, disaster recovery, space flight, law enforcement, scientific investigation, and in-home care, to name a few. A deeper discussion of the opportunities for autonomous systems can be found in a series of recent white papers collected at http://cra.org/ccc/resources/ccc-led-whitepapers/#toward-a-science-of-autonomy-for-physical-systems.

---

[21] Adapted from "Toward a Science of Autonomy for Physical Systems" by Hager, Rus, Kumar, Christensen, accessed at http://cra.org/ccc/wp-content/uploads/sites/2/2015/07/Science-of-Autonomy-June-2015.pdf

We are far from having agents that exhibit the breadth of capabilities described above. Why? At a fundamental level, creating physical intelligence is very hard – what we take for granted, for example carefully grasping the arm of an elderly patient to steady them as they rise from a chair, are fantastically difficulty to engineer. Creating resilient systems that can deal with unforeseen situations and untested failure modes is still an emerging science. Imbuing a system with what we consider "common sense" resists even a clear definition, let alone a robust solution. This doesn't even consider the challenges of communication, instruction, or interaction that we expect from co-workers, co-inhabitants, or others we interact with during the course of a normal day.

Taken together, these technical and socio-technical challenges frame a number of research questions and challenges, each of which is necessary (but perhaps still not sufficient) to achieve the benefits of physical autonomous systems:

**Paths to Autonomy:** How are autonomous systems developed? To what extent is autonomy pre-programmed (innate), versus the results of learning, adaptation, and instruction? How do we imbue these systems with capabilities for self-assessment, self-diagnosis, self-organization, and self-repair?

**Engineering of Autonomy:** Is there a science of integration that can inform the engineering of reliable physically autonomous systems? How does the integration of many sub-systems (as is needed for physically intelligent agents) lead to robust intelligence rather than reliability which decreases as function of the failure modes of each new subsystem. How do we ensure safety?

**Sensing and Autonomy:** How do we translate or adapt new ideas in learning to interpret images, videos, or speech signals into methods to adapt grasping from tactile sensing, to detect and adjust the pose of an object to be placed on a shelf, or to react correctly to the movement of a co-worker? Despite tremendous advances in machine perception, reliable, fast, and robust perception remains a major stumbling block for autonomous systems.

**Autonomy and Human Interaction:** How do we create autonomous systems that are perceived as predictable, reliable and trustworthy? How will we interact with autonomous machines that are ubiquitous in society? How will we communicate our intentions to them, and how will they communicate their intentions to us?

**Autonomy and Society:** What are the policy implications of physical autonomy? What are the societal, legal, and ethical issues? What are the economic implications? How do we frame these issues in ways that do not depend on a specific technology or which become rapidly outdated as science and technology evolve?

### 4.3.2 Future Research Themes

Based on the discussions at the round tables, it is clear that the past five years has moved the field to a new level, which, at the same time, has created new opportunities for fundamental and systems-focused research on new topics and with new capabilities. Some of the themes that emerged during the workshop include the following:

### Learning

*Task-level learning and adaptation:* The current wealth of component capabilities in manipulation, mobility, perception, learning, and reasoning suggest that immensely capable systems should be within grasp. However, the fact is that creating a system to solve complex problems in real-world settings is not "simply" a problem of integrating component systems. For example, most vision modules are developed in isolation from a specific task. Optimization of performance is often based on specific data sets and objectives that may be misaligned from the task, and may in fact be trying to solve a harder problem than necessary to successfully accomplish a specific task. Thus, some type of theory that supports "co-optimization" or "joint evolution" of complex integrated systems will be needed to solve-real world problems.

*Life-long learning:* As robots move from structured, pre-defined tasks to less structured and more variable tasks, it will be incumbent on systems to be able to steadily accumulate experience and adapt their performance to that experience. For example, a construction robot may need to adapt to a different type of building material on each job, or an agriculture robot may optimize its performance as it tills and re-tills the same fields over and over again.

## Software Systems

Safety and Reliability: Should have a high-level supervisor / monitoring process that can help constrain subsystems to validate expected inputs, behaviors, and outputs.

Fault-recovery: better understanding of failure modes / recovery strategies.

Software systems that support rapid and reliable "plug and play" integration of components, but also support adaptation of the resulting systems, and provide guarantees on robustness and resilience of the result. Software components need to become more available in an "app-store" type of context, making it straightforward to download, install, and configure components rapidly.

## Actuation

There are an increasingly myriad of manipulator designs that are flexible, human-safe, and which can be scale and configured for a wide variety of applications. This opens the door to new opportunities to develop highly reconfigurable, integrated, and human-safe systems. For example, prosthetic devices that are "one the fly" customized to the individual, or wearable compliant actuators that provide task-and-person specific augmentations or support. Developing the hardware, control, and software, as well as the integration science to ensure safety, stability, performance, and reliability remain open problems.

A particular subtheme in actuation is soft robotics. Most materials used to build traditional robotic systems are hard materials. As a result, the systems are rigid and bulky. The resulting inertia and the inability of systems to absorb impact makes them unsafe and unsuitable for operation in home and even work environments. In contrast, most of the materials seen in nature are soft. Indeed, there are many new materials such as liquids, foams, and gels, and biological materials that are now being used to develop the next generation of robotic systems. Novel manufacturing techniques also allow us to use these materials to create products, something that was not previously available. While these systems have the potential to be lightweight, deformable, incorporate embedded sensing and actuation, are able to conform to the environment, and can safely interact with humans, they are also difficult to model and harder to control. New approaches to fabrication, modeling, sensing and control will be needed to realize the full potential of soft robotics.

Finally, it is worth noting that employing collections of small, simple robots may soon become a practical reality. Many applications – space, medicine, underwater, or surveillance to name a few – may make use of dozens, hundreds, or thousands of robots (down to the nano scale) to solve problems where access, redundancy, or simply variety are needed.

## Sensors:

Sensing technologies relevant to robotics have continued to improve in price, performance, and resolution. That being said, visual, force, and tactile sensing are still nowhere close to the resolution and sensitivity of the corresponding human senses. In particular, as robotics moves from mobility to manipulation, sensing that supports planning and control of contact and handling of objects will grow to become a major barrier and, therefore by definition, a major research opportunity.

Non-traditional sensing also offers unique opportunities. There are already the first examples of both surface EMG and implanted neural system that offer the disabled the opportunity to regain function they had lost. However, these systems are still in their infancy – we do not understand the transduction, processing, and feedback systems for neural interfaces with a level of fidelity that makes these systems generally usable. Indeed, this is an obvious intersection with the BRAIN initiative which seeks to develop better models for neural systems as part of its charter. Other forms of non-traditional sensing – multispectral imaging, heat, pheromones, galvanic, and so forth offer other opportunities to expand the basis for direct interaction with the environment and with humans.

Sensor architectures are also not yet well developed, in two senses. First, the means of abstracting sensors into task-relevant information is, as yet, a problem-by-problem problem. In order to scale and model sensing in real-world setting, better abstractions that connect sensing to task-relevant and semantically meaningful concepts remain to be developed. Closely related, abstractions for sensors to communicate and combine information are lacking. Work on methods for combining or substituting sensors has continued to make slow progress, but much more remains to be done before sensing can be easily and reliably integrated with actuation, planning, and reasoning in well-understood and well-modeled ways. It is worth noting that uncertainty modeling, often neglected in recently fashionable machine learning methods, is a key need.

## Social Interaction

Robotics is finally entering human environments, from the more structured (roadways, hospitals, nursing homes) to the increasingly less structured (shopping malls, schools, and ultimately homes). Effective co-existence with humans in human environments requires a great deal more than safety and staying out of the way; it requires natural and enjoyable interactions with people on human (not robot) terms. The field of human-robot interaction (HRI), and in particular non-physical, social HRI, is experiencing a major surge in research, development and deployment.

Two major drivers have caused the surge. The first driver is technological, and includes the recent leap in enabling perception technologies through affordable 3D vision for human activity tracking, as well as the development of ever smaller, safer, and, increasingly, softer robot bodies. The second driver is socio-economic, resulting from societal factors (aging population, tech-savvy youth, and safety and health challenges), creating economic opportunities that are causing significant industry investment in robotics development (currently focused on autonomous driving and drones, but expanding into manufacturing and home automation).

HRI contexts vary drastically, from structured ones, such as factories, roadways, airports, and hospitals, to less structured ones, such as streets, public areas, office environments, and retirement homes, to the ultimate unstructured environments: homes. In all cases, HRI involves a combination of real-time perception (of the environment and humans), understanding of not only the current state and ongoing activity, but also intentions of the human participants, and autonomous (or semi-autonomous) response that is safe, timely, natural, ethical, engaging, collaborative, and effective relative to the goals of the interaction context. HRI encompasses one-on-one, one-to-many, and many-to-many human-robot interactions, which span a variety of models for communication and coordination.

Non-physical/social HRI includes the subfields of socially assistive robotics, educational robotics, social robotics, some service robotics, and entertainment robotics. Progress in HRI will require the eventual convergence of the currently separate subareas of physical and social HRI, and more generally, a closer collaboration between robotics, machine vision, machine learning (ML), and AI.

A major barrier in the way of HRI progress is the lack of accessible data sets and evaluation scenarios to ground the work in real world contexts of interest. Because of privacy concerns surrounding the use of human data, and the complexity of deploying robots in real-world human environments, currently very few HRI research projects actually use realistic multi-modal interaction data (featuring audio, video, possibly physiologic data, background data, etc.) and are tested in real-world environments outside of the lab or highly controlled warehouse. It should be noted that robotics in general is in need of more general datasets and scenarios with clear performance metrics.

Research in HRI advanced drastically after the introduction of affordable 3D vision (Kinect, PrimeSense) and the associated models of human activity, facilitating recognition and tracking needed for HRI. As outlined above, similar leaps in capability could be achieved by removing some of the barriers, including providing training data sets, evaluation testbeds and environments, and synergies with machine vision and machine learning research.

The following are some of the challenging areas of non-physical / social HRI research in the coming years.

Degrees of autonomy: As with any intelligent system, the level of autonomy vs. user control is important, but it becomes particularly interesting when the system is socially engaging and potentially persuasive and involved in the user's daily and social life. Determining natural and appropriate ways for the user to determine and adjust the autonomy of the system in real time presents interesting and novel research challenges.

Enjoyment of interaction: The vast majority of robotics to date has focused on functional systems, but social HRI aims for user engagement and enjoyment. To achieve this, synergies with social scientists as well as creative interaction designers (such as developers of movie and video game characters) is necessary and needs to be facilitated. It also needs to be treated with proper care since both unwanted attachment and unmet expectations constitute undesirable outcomes of the technology.

Privacy and security: The general challenges in data privacy and network security are at their peak with the type of personal and sensitive information obtained from face-to-face video and audio interactions with people, including children and special needs users among other vulnerable populations. Proactively focusing attention on proper treatment of these issues is important or public backlash from early failures may cost the field significant delays.

Trust and awareness: Beyond privacy and security, the issue of trust between the user and robot is one of the most sensitive. Establishing trust is already an established research topic in AI and simulated agents, but in the context of socially aware machines, the challenge may be less about establishing trust and more about managing it properly and ethically.

Robotics in health and wellness: The role of social and socially assistive robots in human health and wellness in a variety of settings, from managed care (retirement homes, nursing homes, hospitals, etc.) to in-home care, is an area expected to grow quickly due to the vast need and gap in available human resources. A great many research challenges remain in order to design machines that can assess actual human needs in real time and provide appropriate, personalized, ethical, and timely feedback, companionship, and care. Currently the focus of discussion is on care for the growing elderly population, but the span of technology needs and niches ranges from the very young to the very old, along with a broad range of user capabilities and needs (cognitive, physical, and emotional), creating numerous research challenges for the field and for interdisciplinary collaborations well beyond the field itself in order to make significant impact beyond the lab and into real world use.

# Enabling infrastructure

Robotics technology has never been more accessible. Arms, hands, and software are cheaper and more capable than ever before. However, many of the application spaces for robotics demand substantial infrastructure – hardware, software and data. For example:

1) Automated driving requires cars, areas to drive, and instrumentation to test and measure systems responses. Many major automotive companies (in the US, as least GE, Ford, and Toyota) are putting this infrastructure in place. However, it is not yet clear how open these platforms will become, and thus how much the academic research community will be able to participate in these developments.

2) Advanced collaborative manufacturing often requires realistic factory conditions and deep understanding of the real-world problems of deploying systems. Currently, researchers largely seek out and form their own collaborative relationships with companies. However, this makes it difficult to test and compare competing approaches, and understand and improve on system performance in a standardized manner.

3) Medical robotics requires substantial collaboration, and expensive and unique commercial platforms upon which research can build. Very few groups can carry a research project from the lab into the OR, and doing so when approved, capable platforms exist wastes resources and energy better devoted to new innovation.

4) Data and cloud capabilities are beginning to crop up as a trend in robotics. For example, Google-X can afford to create a "robot farm" and use that farm to "harvest" data on e.g. manipulation of objects. No academic group, on its own, can afford to undertake a similar effort, though early crowd-sourcing efforts are underway. Models that encourage both the academic and the industry community to share data will become ever more essential to progress.

As the field moves forward, understanding and creating incentives and modes of access to shared research infrastructure will both allow a broader range of individuals to participate in robotics research, and will serve to better standardize and quantify measures of progress for the field.

# Wearable Embedded Devices

The NRI had, as a large driver, co-robotic systems, i.e., robotic systems that interact synergistically with humans. Yet the focus on humans interacting with robots could lead to new challenges that extend beyond the current NRI program. Of special note is that wearable devices, as a focused area, would require a depth of understanding in soft robotics, including but not limited to novel materials, actuators, control and sensing, nonconventional substrates and a direct connection to biology and bio-inspired models. These devices could be worn by human users, and indeed embedded in human users, and therefore extend well beyond robotics. The importance of this area could be far-reaching in the context of application domains ranging from the medical domain, e.g., rehabilitation, to use by millions of Americans in their daily lives.

# Collaborative Systems

Many processes are becoming more and more human centered. Humans play a key role in the management of ever increasing complexity, for processes that require significant cognitive reasoning and rapid evolution in product definition or mix.

In the future we will utilize multi modal interfaces, intuitive and user experience driven work×flows, to safely plan, program, operate, and maintain manufacturing systems. Mobile and ubiquitous technology will allow workers to remotely control and supervise manufacturing operations. New safety systems will allow full adaptation of worker–robot collaboration that will enhance competitiveness and compensate for age- or inexperience related worker limitations. Dynamic reallocation of tasks and changes in automation levels will enable human–automation symbiosis and full deployment of the skills of the workforce. Enhancement and support of the workers' cognitive skills will become increasingly important to create human centered workplaces.

Human-machine interaction has evolved significantly through new and emerging safety standards such as ISO 10218.6 and R15.06. The clear definition of models and methods for interaction allows design of systems at a much lower cost and with improved performance as seen for collaborative robotics. A major challenge is the need for application specific safety certification.

## 4.4 Educational Opportunities

Robotics is a universal educational vehicle. As noted above, at the graduate level, more and more universities are setting up graduate programs that include a core educational component as well as research training. The growth in the major robotics conferences is reflective of the growth in student interest in the field.

Graduate programs are providing students that have broad knowledge across control, artificial intelligence, autonomy, mechanics, perception and human-interaction. The programs include both coursework and multidisciplinary research with the objectives of:

a. Educating students in the engineering and science principles necessary to generate novel perspectives, concepts, and technologies required to push the boundaries of robotics;

b. Instill the desire to pursue life-long learning;

c. Conduct fundamental and applied research in the domains necessary to create new knowledge and technologies that have high societal and economic impact;

d. Produce graduates who have the necessary skill sets to rise to leadership positions in academia and industry.

A number of M.Sc. and Ph.D. programs are in place across the country. More recently a number of undergraduate programs or minors have also emerged. There is a strong demand from industry for graduates from these programs. Clearly it would be desirable to consider ways of coordinating some of these educational programs.

For training of robot operators there are relatively few example of broader programs that consider this. One such program is the RAMTEC[22] in Ohio which provide operator training across many different industry providers. The program by TAMU on disaster management have similar program for disaster scenarios. To our knowledge there are no general programs at the certificate level and there is a clear need across US for such training.

At the STEM level the two biggest programs are US FIRST[23] and BEST[24]. Jointly these programs reach more than 100,000 students each year. There is a significant opportunity to leverage programs like these to promote STEM education. One challenge for some of these effort has been in the outreach to minority communities. The cost of participation can sometimes be prohibitive.

---

[22] http://www.ramtecohio.com
[23] http://www.firstinspires.org
[24] http://www.bestinc.org

## 4.5 Community Building

The CPS community has been very successful at organizing a virtual organization[25] that manages annual meetings, a highly successful web facility for broader outreach, and coordination of a roadmap process engaging both academia and industry.

Within the robotics community annual meetings have been organized but without a clear "community" organization. An embryo for a CPS like organization – the robotics-vo was launched 2012, but it so far not managed to become self-sustainable. The present Robotics-Vo web facility is sponsored by private funds[26] and may not be sustainable. There is a clear need for an organization that manages annual meetings, a regular road-mapping process and general dissemination. It is not at this time clear how such an organization may be financed.

## 5. Summary

The past five years have seen enormous strides in both the fundamental research and the applications of robotics technologies. Robotics is well on its way to being firmly established as an academic discipline as well as a potent force in future technology commercialization. Many technologies have been "democratized," meaning they are now far more widely accessible, driving a much broader space of activities and opportunities in education, research, and technology transfer. Robotics creates excitement in nearly every population it touches.

However, if we use the auto industry as an analogy, it is not far off to consider robotics today to be still just out of the "Henry Ford" stage. Today's technologies are really just the first platforms upon which future innovations will be built. The current NRI has helped to explore the possible spectrum of robotics applications, and it has, in particular, introduces human-robot interaction as a first class concept in the field. However, it has also posed new problems and barriers, many of which have been discussed above.

Taking the next steps toward the future relies on fundamental research across many disciplines, as well at the integrative science to draw that work together. As robots move from highly structured environments, and begin to interact with the real world, we foresee barriers that current methods and technologies cannot overcome. As the application space scales, the need for better sensing, better actuation, and more general planning, reasoning, and learning will become paramount. The ability to rapidly architect, implement, deploy, and adapt new systems to new problems will require new concepts in software, and new methods of integration.

If we continue on this path, it seems clear that robotics will create an entire new sector of the economy. NRI and similar programs are thus providing an on-ramp into a new set of educational and economic opportunities for the nation that we cannot ignore if we hope to continue to lead world innovation in new technologies.

## Acknowledgement

The CCC support for this study is gratefully acknowledged.

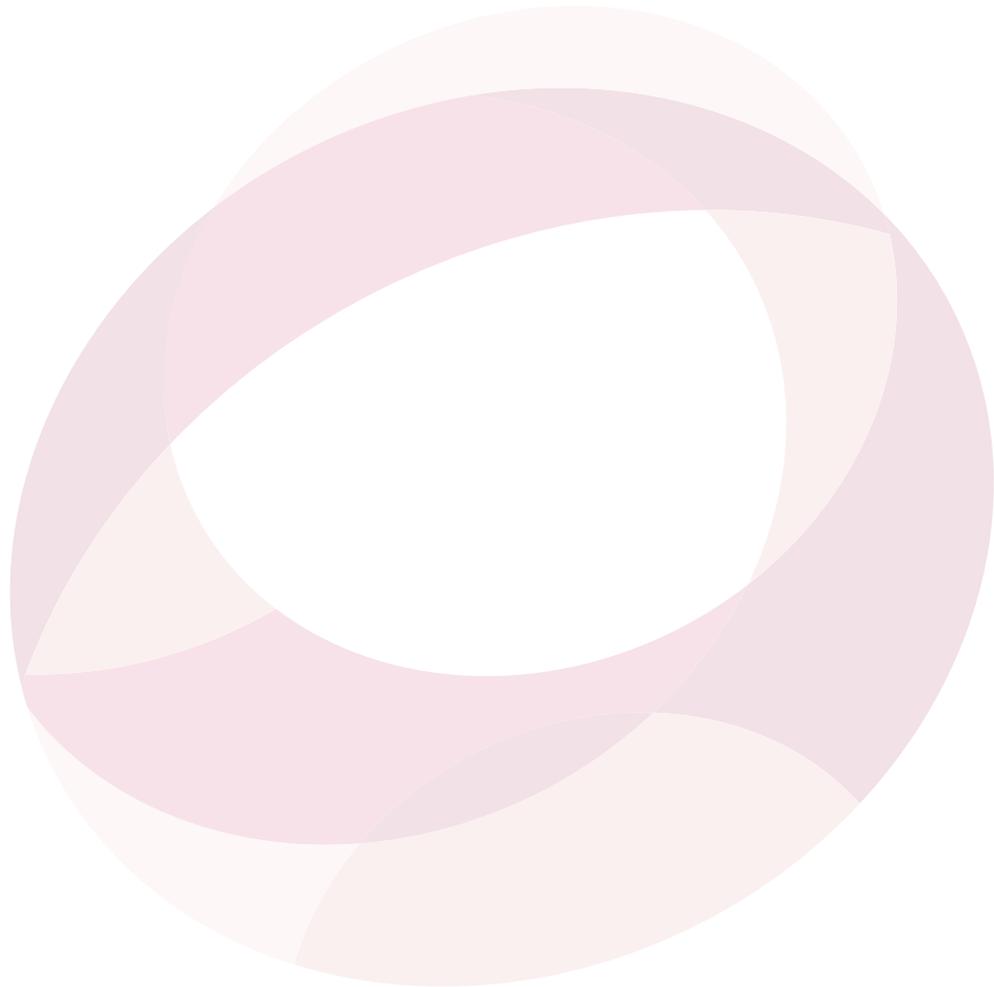

This material is based upon work supported by the National Science Foundation under Grant No. 1136993. Any opinions, findings, and conclusions or recommendations expressed in this material are those of the authors and do not necessarily reflect the views of the National Science Foundation.